\documentclass[prl,twocolumn,groupedaddress,floatfix]{revtex4-1}

\usepackage{lipsum} 
\usepackage{verbatim}
\usepackage{epsfig}
\usepackage{subfigure}
\usepackage{graphicx}
\usepackage{amsfonts}
\usepackage[figuresright]{rotating}
\usepackage{amssymb}
\usepackage{amsmath}
\usepackage{psfrag}
\usepackage{esint} 
\usepackage{bm}
\usepackage[colorlinks,linkcolor=blue,anchorcolor=blue,citecolor=blue,urlcolor=blue]{hyperref}
\usepackage[version=4]{mhchem}

\usepackage{diagbox}

\def\be{\begin{equation}} \def\ee{\end{equation}}
\def\bea{\begin{eqnarray}} \def\eea{\end{eqnarray}}

\renewcommand{\vec}[1]{\mathbf{#1}}
\newcommand{\vecg}[1]{\boldsymbol{#1}}

\newcommand{\ket}[1]{| #1 \rangle}
\newcommand{\bra}[1]{\langle #1 |}
\newcommand{\braket}[2]{\langle #1 |#2\rangle}

\newcommand{\acomm}[2]{\left\{#1,#2\right\}}
\def\bpm{\begin{pmatrix}} \def\epm{\end{pmatrix}}

\renewcommand{\vr}{\mathbf{r}}
\newcommand{\vk}{\mathbf{k}}

\newcommand{\vq}{\mathbf{q}}
\newcommand{\vqp}{{\mathbf{q}^\prime}}

\newcommand{\FS}{\text{FS}}
\newcommand{\vd}{\mathbf{d}}
\renewcommand{\d}{\text{d}}
\newcommand{\expect}[1]{\langle #1 \rangle}

\DeclareMathOperator{\SU}{SU}
\DeclareMathOperator{\SO}{SO}
\DeclareMathOperator{\Un}{U}
\DeclareMathOperator{\Tr}{Tr}
\DeclareMathOperator{\Pf}{Pf}
\DeclareMathOperator{\sgn}{sgn}
\DeclareMathOperator{\diag}{diag}
\DeclareMathOperator{\Rea}{Re}
\makeatletter
\newcommand*{\balancecolsandclearpage}{%
  \close@column@grid
  \clearpage
}
\makeatother

\begin{document}
\title{$\mathbb{Z}_2$ Topologically Obstructed Superconducting Order}
\author{Canon Sun}
\author{Yi Li}
\address{Institute for Quantum Matter and Department of Physics and Astronomy, Johns Hopkins University, Baltimore, Maryland 21218, USA}
\date{\today}

\begin{abstract}
We propose a class of topological superconductivity in which the pairing order is $\mathbb{Z}_2$ topologically obstructed in a three-dimensional time-reversal invariant system. When two Fermi surfaces are related by time-reversal and mirror symmetries, such as those in a $\mathbb{Z}_2$ Dirac semimetal, the inter-Fermi-surface pairing in the weak-coupling regime 
inherits the band topological obstruction. As a result, the pairing order cannot be well-defined over the entire Fermi surface and forms a time-reversal invariant generalization of U($1$) monopole harmonic pairing. A tight-binding model of the $\mathbb{Z}_2$ topologically obstructed superconductor is constructed based on a doped $\mathbb{Z}_2$ Dirac semimetal and exhibits nodal pairings. At an open boundary, the system exhibits a time-reversal pair of topologically protected surface states. 
\end{abstract}

\maketitle

{\it Introduction.} -- 
Central to understanding the properties of a superconductor is the symmetry of its pairing order, which forms irreducible representations of the symmetry of the system, and is usually characterized by the spherical harmonic functions or their lattice counterparts.  
Notable examples include conventional $s$-wave superconductors such as \ce{Hg} and \ce{Nb}, 
unconventional $p$-wave superfluid  ${}^3\ce{He}$  \cite{Anderson1961,Balian1963,Leggett1975,Volovik2009}, $p$-wave heavy fermion compounds \cite{Stewart1984,Sigrist1991,Pfleiderer2009},
and $d$-wave high-$T_c$ cuprates \cite{Harlingen1995,Tsuei2000}. 

Another notion fundamental to unearthing new phases of matter is the topology of electronic bands. 
In a seminal work, a two-dimensional (2D) insulating system with broken time-reversal symmetry (TRS) has been discovered to exhibit the quantum anomalous Hall effect characterized by a non-zero Chern number \cite{Haldane1988}. It arises from the geometry of Bloch wave functions whose phase cannot be well defined over the entire 2D Brillouin zone (BZ) \cite{Kohmoto1985}. 
The notion of topology in electronic bands was then generalized to insulators with TRS in two and three dimensions (3D), which are characterized by $\mathbb{Z}_2$ invariants \cite{Kane2005a,Bernevig2006a,Fu2007,Fu2007a,Roy2009,Qi2008,Moore2007,Roy2009a,Lee2008,Soluyanov2011}.
Furthermore, topological obstructions in metallic bands and quasiparticle states give rise to the notions of topological Fermi liquids, semimetals and superconductors \cite{Haldane2004, Murakami2007, Wan2011,Burkov2011,Wang2012,Hosur2013, Borisenko2014, Xu2015, Xiong2015a, Moll2016, Bansil2016, Armitage2018, Volovik2009,Read2000,Kitaev2001,Fu2008,Qi2009a,Lutchyn2010,Qi2011,Ando2015,Sato2017,Schindler2020}.

A recent work introduced the notion of \emph{monopole superconductivity} \cite {Li2018}, which captures an $\Un(1)$ topological obstruction in the phase of the superconducting order \cite{Murakami2003a}. In contrast to traditional discussions of topological superconductivity, which generally revolve around the topology of Bogoliubov-de Gennes (BdG) quasiparticle states, in a monopole superconductor it is the pairing order, a direct physical observable, that is topologically obstructed. This obstruction leads to nodal superconducting gap functions described by the monopole harmonics \cite{Li2018}, which are topological sections of an $\Un(1)$ bundle over a sphere \cite{Wu1976,Wu1977}. 
This monopole pairing is fundamentally different from the familiar $s$-, $p$- and $d$-wave pairings based on the spherical harmonics and is beyond the ten-fold way classification \cite{Schnyder2008}. 
It can be realized in certain doped Weyl semimetals or spin-orbit coupled cold atom systems \cite{Li2018,Sun2019,Li2020,Muoz2020}. The notion of monopole pairing has also been extended to the particle-hole channel and leads to, for example, monopole density wave orders \cite{Bobrow2019}.


In this letter, we explore a non-Abelian topological obstruction in superconducting pairing orders characterized by a $\mathbb{Z}_2$ invariant. 
When two helical FSs are related by time-reversal (TR) and mirror symmetries with their composed symmetry $\hat{\mathcal{T}}$ satisfying $\hat{\mathcal{T}}^2=-1$, 
the Bloch states at the FSs are classified by a $\mathbb{Z}_2$ topological index. 
In the weak-coupling regime, when Cooper pairing occurs between $\mathbb{Z}_2$ non-trivial FSs, the topology of Bloch states near FSs induces an $\SO(3)$ topological obstruction in the superconducting order which is characterized by a $\mathbb{Z}_2$ invariant. 
This obstruction is the inability to enforce the symmetry condition imposed by $\hat{\mathcal{T}}$ on the pairing order globally without introducing singularities, which can be made regular by relaxing the symmetry condition through the introduction of the sewing matrix.
An example of $\mathbb{Z}_2$-obstructed pairing is explored in a tight-binding model of a doped $\mathbb{Z}_2$ Dirac semimetal in proximity to an $s$-wave superconductor with inter-orbital pairing. The system exhibits a time-reversal pair of topological surface states which form zero-energy Majorana arcs connecting the surface projections of bulk gap nodes.

\begin{figure}[tpb]
{\centering  {\epsfig{file=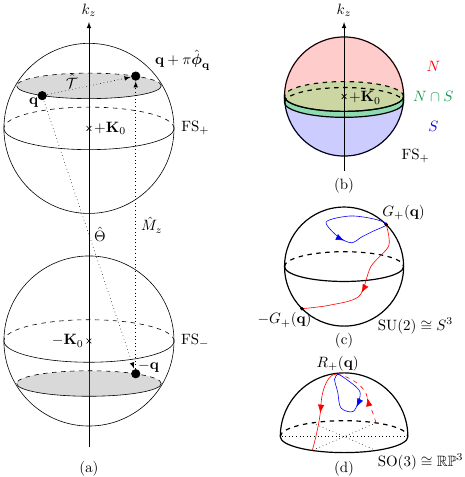,width=0.75\linewidth}}
}
\caption{(a) Two Fermi surfaces, $\FS_{\pm}$, enclosing  $\pm \vec{K}_0$ located along the $k_z$-axis, are related by $\hat{\Theta}$, $\hat{M}_z$, and $\hat{\mathcal{T}}$ symmetries. (b) The 2D spherical FS is divided into two gauge patches $N$ (red) and $S$ (blue), which overlap at the equator, $N\cap S$ (green). (c) From Ref. \onlinecite{Lee2008}, the two topological classes of $\SU(2)$ transition matrices in cases (I) and (II) correspond to the contractible, blue and non-contractible, red loops, respectively, in $\SU(2)\cong S^3$. (d) The group manifold of $\SO(3)$ is $\mathbb{RP}^3$. Under the map $G_+(\vq)\mapsto R_+(\vq)$, the two paths in (c) are mapped to paths of the corresponding color in (d).
}
\label{fig: spheres}
\end{figure}

{\it Topological $\mathbb{Z}_2$ Fermi Surfaces.} -- 
We begin with a 3D minimal model of a pair of $\mathbb{Z}_2$ obstructed FSs. Consider two disjoint spherical Fermi surfaces, FS$_\pm$, centered about $\pm \vec{K}_0=\pm(0,0,K_0)^T$ related by TR and `mirror' symmetries,  
as shown in Fig. \ref{fig: spheres} (a). 
Let $\hat{\Psi}^\dagger_{\pm,a}(\vq)$ denote the fermion creation operator on FS$_{\pm}$ for the $a$-th band with wavevector $\vk=\pm \vec{K}_0+ \vq$. Here $a=1,2$ labels the pseudospin 
degrees of freedom on the FSs. 
The fermion operators on FS$_+$ are related to those on FS$_-$ by TRS, $\hat{\Theta}$, as  
$\hat{\Theta}\hat{\Psi}^\dagger_{\pm,a}(\vq)\hat{\Theta}^{-1}
=\sum_b\hat{\Psi}^\dagger_{\mp,b}(-\vq)[i\sigma_y]_{ba}$, 
where $\theta_\vq$ and $\phi_\vq$ are the polar and azimuthal angles of $\vq$, respectively. 
If the `mirror' symmetry, $\hat{M}_z$, satisfies   $\hat{M}_z\hat{\Psi}^\dagger_{\pm,a}(\theta_\vq,\phi_\vq)\hat{M}_z^{-1}
= \hat{\Psi}^\dagger_{\mp,a}(\pi-\theta_\vq,\phi_\vq)$, 
the combination $\hat{\mathcal{T}}\equiv \hat{M}_z \hat{\Theta}$ leads to a new antiunitary symmetry that relates operators on the same FS at the same $k_z$,
\begin{equation}\label{eq: TR FS}
\hat{\mathcal{T}}\hat{\Psi}^\dagger_{\pm,a}(\vq)\hat{\mathcal{T}}^{-1}=\sum_b\hat{\Psi}^\dagger_{\pm,b}(\vq+\pi \hat{\vecg{\phi}}_\vq)[i\sigma_y]_{ba},
\end{equation}
satisfying $\hat{\mathcal{T}}^2=-1$. This can be viewed as TRS in 2D.
 
Based on the analysis of physical TRS in 2D topological insulators in Refs. \onlinecite{Lee2008,Roy2009a}, the symmetry $\hat{\mathcal{T}}$ here classifies the Bloch states near a FS into two topological sectors. 
To illustrate this, let us first focus on $\FS_+$ and divide it into two patches, $N$ and $S$, as shown in Fig. \ref{fig: spheres} (b). Define the fermion operators $\hat{\Psi}^\dagger_{+,a}(\vq)$ in $N$ as  $\hat{\alpha}^\dagger_{+,a}(\vq)$ and those in $S$ as $\hat{\beta}^\dagger_{+,a}(\vq)$ with their respective Bloch states smoothly defined over each patch. 
Generally, the operators defined on the two patches are related by an $\Un(2)$ gauge transformation $M_{+}(\vq)$ 
at the overlap, 
\begin{align}\label{eq: gauge transformation}
    \hat{\alpha}^\dagger_{+,a}(\vq)&=\sum_{b=1}^2\hat{\beta}^\dagger_{+,b}(\vq)M_{+,ba}(\vq),&  \vq &\in N\cap S.
\end{align}
Here, 
the transition matrix 
$M_{+}(\vq)=e^{i\omega_+(\vq)}G_+(\vq)$ consists of an $\Un(1)$ phase $e^{i\omega_+(\vq)}$ and $\SU(2)$ matrix
$G_{+}(\vq)$. 

In order for the transition matrix to be compatible with $\hat{\mathcal{T}}$-symmetry, it is constrained to one of two possible cases \cite{Lee2008,Roy2009a}: (I) $G_{+}(\vq+\pi\hat{\vecg{\phi}}_\vq)=G_{+}(\vq)$ and (II) $G_{+}(\vq+\pi\hat{\vecg{\phi}}_\vq)=-G_{+}(\vq)$.
$G_{+}$ maps 
$\vq \in N\cap S \cong S^1$ to 
$\SU(2)$ matrices. When 
$\vq$ is varied from $(\theta_\vq,\phi_\vq)$ to $(\theta_\vq,\phi_\vq+\pi)$ 
along $S^1$, $G_+$ traces out a path in $\SU(2)$. As shown in  Fig. \ref{fig: spheres} (c), the path  in $\SU(2)$ is contractible in case (I)
which implies the transition matrix can be continuously deformed to the identity, and hence the Bloch states can be smoothly defined over FS$_+$. 
On the other hand, in case (II), there does not exist such a deformation as the path must visit the antipodal point. This topological obstruction makes it necessary to define the Bloch states using two gauge patches. 

The operators on $\FS_-$ are in the same topological class as those on $\FS_+$. This follows from TRS, which relates the transition matrix on $\FS_{-}$, $M_-(-\vq)$, with $M_+(\vq)$ via
\begin{equation}\label{eq: M+ and M-}
    M_-(-\vq)
    =e^{-i\omega_+(\vq)}G_+(\vq).
\end{equation} 
As the transition matrices $M_\pm$ share the same $\SU(2)$ part, states on FS$_\pm$ fall the same topological class.

It is possible to choose Bloch states that are globally well-defined  
if the $\hat{\mathcal{T}}$ condition is not strictly enforced \cite{Lee2008,Fu2006}. 
Let  $\hat{\Psi}^\dagger_{\pm,a}(\vq)=\hat{\chi}^\dagger_{\pm,a}(\vq)$ denote the creation operator for a state that is regular over the entire FS. The condition imposed by $\hat{\mathcal{T}}$-symmetry is relaxed to
\begin{equation}\label{eq: TR sewing}
    \hat{\mathcal{T}}\hat{\chi}^\dagger_{\pm,a}(\vq)\hat{\mathcal{T}}^{-1}=\sum_b \hat{\chi}_{\pm,b}^\dagger(\vq+\pi \hat{\vecg{\phi}}_\vq)w_{\pm,ba}(\vq).
\end{equation}
Here $w_{\pm}(\vq)\equiv \bra{0}\hat{\chi}_{\pm}(\vq+\pi \hat{\vecg{\phi}}_\vq) \hat{\mathcal{T}}\hat{\chi}_{\pm}^\dagger(\vq)\ket{0}$ are unitary matrices defined on FS$_{\pm}$ called sewing matrices. Since they are not independent and satisfy $w_{\pm}(\theta_\vq,\phi_\vq ) =w_{\mp}(\pi-\theta_\vq,\phi_\vq)$ as a result of mirror symmetry,   
let us focus on $w_+(\vq)$. Because $\hat{\mathcal{T}}^2=-1$, the sewing matrix satisfies $w_{+}(\vq)=-w^T_{+}(\vq+\pi\hat{\vecg{\phi}}_\vq)$, which makes it antisymmetric at the $\hat{\mathcal{T}}$-invariant momenta, namely the north and south poles ($\theta_\vq=0,\pi$). As an unitary matrix, it can be decomposed as $w_{+}(\vq)=e^{i\zeta_{+}(\vq)}\tilde{w}_{+}(\vq)$, where $e^{i\zeta_{+}(\vq)}\in \Un(1)$ and $\tilde{w}_{+}(\vq)\in\SU(2)$. At the two poles, $\tilde{w}_+(\theta_\vq=0,\pi)=\Pf\tilde{w}_+(\theta_\vq=0,\pi) i\sigma_y$ and the $\mathbb{Z}_2$ invariant of $\FS_+$ has been defined using the Fu-Kane formula \cite{Fang2016}
\begin{equation}\label{eq: Pfaffian}
    \delta=\Pf \tilde{w}_+(\theta_\vq=0)\Pf \tilde{w}_+(\theta_\vq=\pi),
\end{equation}
which takes value $+1$ $(-1)$ in the non-topological (topological) phase. When inversion symmetry, defined as $
\hat{\Pi}\hat{\chi}^\dagger_{\pm,a}(\vq)\hat{\Pi}^{-1}=\sum_b \hat{\chi}^\dagger_{\mp,b}(-\vq)U_P$, where $U_P$ is unitary,
is present, $\delta$ reduces to the product of the eigenvalues of the in-plane inversion operator, defined as $\hat{P}\equiv \hat{M}_z\hat{\Pi}$, at the two $\hat{\mathcal{T}}$-invariant points \cite{Fu2007, Fang2016}. The equivalence between the two gauges picture and the Fu-Kane invariant is established in Supplementary Material (S.M.) I.

{\it Topologically Obstructed Superconducting Order.} --  The $\mathbb{Z}_2$ obstructed FSs can induce a topological obstruction in the pairing order. Let us consider inter-FS Cooper pairing between FS$_+$ and FS$_{-}$, as described by the mean-field pairing Hamiltonian
\begin{equation}\label{eq: MF Hamiltonian}
    \hat{H}_{\Delta}=\sum_{\vq,a,b}\hat{\alpha}^\dagger_{+,a}(\vq)\Delta_{ab}^N(\vq)\hat{\alpha}^\dagger_{-,b}(-\vq)+\text{h.c.},
\end{equation}
where $\Delta^N_{ab}(\vq)=\sum_{\vqp,cd} V_{abcd}(\vq,\vqp)\expect{\hat{\alpha}_{-,c}(-\vqp)\hat{\alpha}_{+,d}(\vqp)}$ is the superconducting pairing matrix defined in the region $N$ of $\FS_+$ and $V_{abcd}(\vq,\vqp)$ is the inter-FS attractive interaction potential. To obtain the pairing matrix in the region $S$, we perform the gauge transformation in Eq. \eqref{eq: gauge transformation}, leading to the relation $\Delta^S(\vq)=G_+(\vq)\Delta^N(\vq)G^T_+(\vq)$, where we have used Eq. \eqref{eq: M+ and M-}. It is convenient to decompose the pairing matrix into singlet and triplet sectors, $\Delta^{N/S}(\vq)=\left(d_0^{N/S}(\vq)+\vec{d}^{N/S}(\vq)\cdot \vecg{\sigma}\right)i\sigma_y$. In this notation, the pairing matrix transforms under the gauge transformation as
\begin{align}
d_0^S(\vq)&=d_0^N(\vq)\\
\vec{d}^S(\vq)\cdot \vecg{\sigma}&=G_+(\vq)\left(\vec{d}^N(\vq)\cdot \vecg{\sigma}\right)G_+^\dagger(\vq)\nonumber\\
&=\left(R_+(\vq)\vec{d}^N(\vq)\right)\cdot\vecg{\sigma},
\end{align}
where $R_+(\vq)$ is the rotation matrix in the vector representation associated 
with $G_+(\vq)$. The effect of the gauge transformation is a rotation on the spin-$0$ and spin-1 sectors of the pairing matrix. The singlet sector $d_0^N$ is unaffected by the gauge transformation as it is rotationally invariant. Therefore, for singlet pairing in the band basis the pairing matrix can be smoothly defined over the FS. On the other hand, $\vec{d}^N(\vq)$ is generally not invariant under the rotation, with the only exception being when $\vec{d}^N(\vq)$ is parallel to the rotation axis. The map $G_+(\vq)\mapsto R_+(\vq)$ transforms the two classes of paths in $\SU(2)$ space to loops in $\SO(3)$. As illustrated in Figs. \ref{fig: spheres} (c) and (d), a path belonging to case (I) is mapped to a contractible loop whereas one in case (II) is mapped to a non-contractible loop, corresponding to the trivial and non-trivial elements of the fundamental group $\pi_1(\SO(3))\cong \mathbb{Z}_2$, respectively.  Therefore, if the FSs are $\mathbb{Z}_2$ non-trivial, the superconducting order parameter associated with triplet pairing is topologically obstructed and requires the use of two gauge patches.

In the sewing matrix approach, it is possible to select a single gauge patch to describe the pairing matrix, at the expense of the $\hat{\mathcal{T}}$-symmetry condition on the $\vec{d}$ vectors. In the singular gauge, the symmetry $\hat{\mathcal{T}}$ imposes on the pairing matrix the constraint $(i\sigma_y)\left[\Delta^{N/S}(\vq)\right]^*(i\sigma_y)^T=\Delta^{N/S}(\vq+\pi\hat{\vecg{\phi}}_\vq)$. 
For singlet pairing, this gives  $\left[d_0^{N/S}(\vq)\right]^*=d_0^{N/S}(\vq+\pi \hat{\vecg{\phi}}_\vq)$ and for triplet pairing $-\left[\vec{d}^{N/S}(\vq)\right]^*=\vec{d}^{N/S}(\vq+\pi \hat{\vecg{\phi}}_\vq)$. To study what the $\hat{\mathcal{T}}$-symmetry condition is in the regular gauge, first perform a basis transformation from $\hat{\alpha}^\dagger_{\pm}(\vq)$ to $\hat{\chi}^\dagger_{\pm}(\vq)$. The pairing matrix in the non-singular basis reads $\Delta(\vq)=U_+(\vq)\Delta^N(\vq) U_-^T(-\vq)$, where $[U_{\pm}(\vq)]_{ab}=\bra{0}\hat{\chi}_{\pm,a}(\vq) \hat{\alpha}^\dagger_{\pm,b}(\vq) \ket{0}$. Using mirror and $\hat{\mathcal{T}}$ symmetries, the basis transformation matrices are related by $U_-^T(-\vq)=(-i\sigma_y)U_+^\dagger(\vq)w_+(\vq+\pi \hat{\vecg{\phi}}_\vq)$, and hence
\begin{equation}\label{eq: gap sewing}
    \Delta(\vq)=\left( d_0(\vq)+\vd(\vq)\cdot \vecg{\sigma}\right)w_+(\vq+\pi \hat{\vecg{\phi}}_\vq),
\end{equation} where $d_0=d_0^N$, $\vd(\vq)=D_+(\vq)\vec{d}^N(\vq)$, and $D_+(\vq)$ is the $\SO(3)$ rotation matrix associated with the $\SU(2)$ part of $U_+(\vq)$.  In this new basis, the condition imposed by $\hat{\mathcal{T}}$-symmetry is $w_+(\vq)\Delta^*(\vq)w^T_+(\vq+\pi \hat{\vecg{\phi}}_\vq)=\Delta(\vq+\pi \hat{\vecg{\phi}}_\vq)$. For singlet pairing, this simplifies to the same condition as in the singular gauge and thus there is no obstruction to the $\hat{\mathcal{T}}$-symmetry condition. In contrast, for triplet pairing, $w_+(\vq)\left(\vd(\vq)\cdot\vecg{\sigma} \right)^*w^\dagger_+(\vq)=\vd(\vq+\pi \hat{\vecg{\phi}}_\vq)\cdot \vecg{\sigma}$. This cannot be reduced to the ordinary time-reversal condition since $\tilde{w}_+(\theta_\vq=0)=-\tilde{w}_+(\theta_\vq=\pi)$ for a $\mathbb{Z}_2$ non-trivial FS. Consequently, there is an obstruction to enforce the $\hat{\mathcal{T}}$-symmetry condition for triplet pairing, in agreement with the result obtained in the singular gauge.

Apart from TRS, 
mirror symmetry in the fermion BdG Hamiltonian 
requires the pairing matrix to satisfy
$\Delta^{N/S}(\vq)=-[\Delta^{N/S}(\vq+\pi \hat{\vecg{\phi}}_\vq)]^T$. This implies the pairing matrix in the triplet channel vanishes at the $\hat{\mathcal{T}}$-invariant points. In other words, the resulting BdG quasiparticle spectrum is nodal at the poles.

{\it A Model of the $\mathbb{Z}_2$ Pairing Order.} 
-- A simple example of  $\mathbb{Z}_2$-obstructed pairing can be constructed by considering inter-FS Cooper pairing in a $\mathbb{Z}_2$ Dirac semimetal \cite{Morimoto2014}: 
$\hat{H}=\hat{H}_0+\hat{H}_{\Delta}$, where  $\hat{H}_0=\sum_{\vk,\alpha\beta}\hat{c}^\dagger_{\vk\alpha}(h(\vk)-\mu)_{\alpha\beta}\hat{c}_{\vk\beta}$ is a four-band Hamiltonian of a  $\mathbb{Z}_2$ Dirac semi-metal and  $\hat{H}_\Delta=\sum_{\vk,\alpha\beta}\hat{c}^\dagger_{\vk\alpha}\bar{\Delta}_{\alpha\beta}(\vk)\hat{c}^\dagger_{-\vk\beta}+\text{h.c.}$
describes the proximity-induced mean-field inter-FS pairing. Here, the chemical potential $\mu>0$, $\bar{\Delta}_{\alpha\beta}(\vk)$ is the pairing matrix, and $\hat{c}^\dagger_{\vk\alpha}$ is the creation operator for an electron with wavevector $\vk$ and index $\alpha$, which labels the orbital and spin degrees of freedom.

\begin{figure}[tpb]
\subfigure[]{\centering  {\epsfig{file=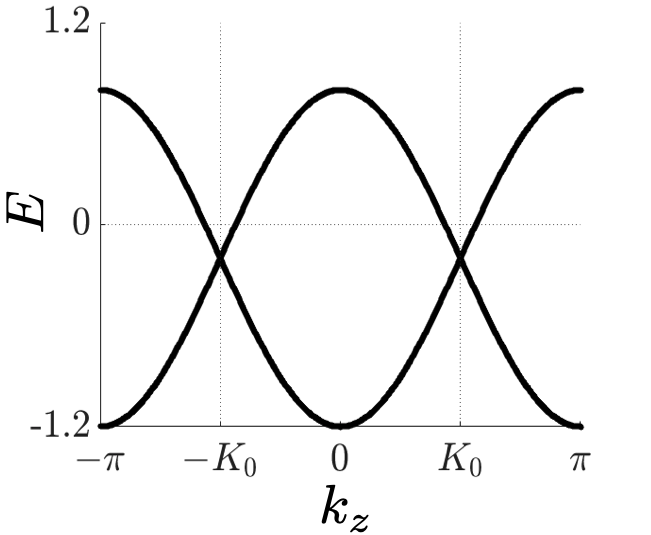,width=0.45\linewidth}}
}
\subfigure[]{\centering  {\epsfig{file=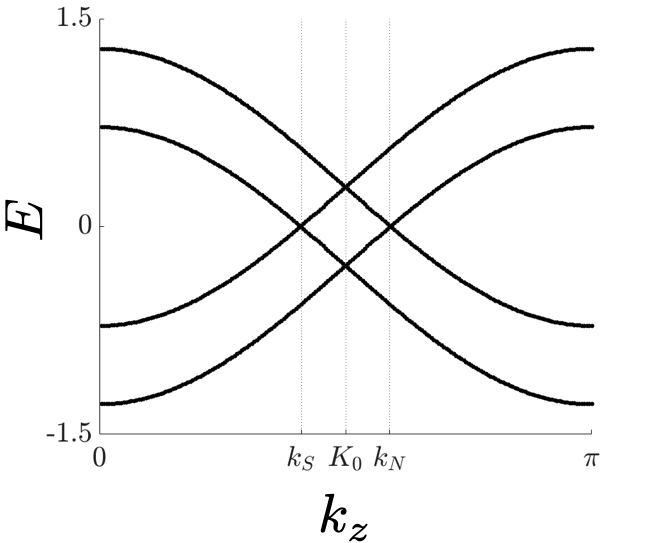,width=0.45\linewidth}}
}
\subfigure[]{\centering \epsfig{file=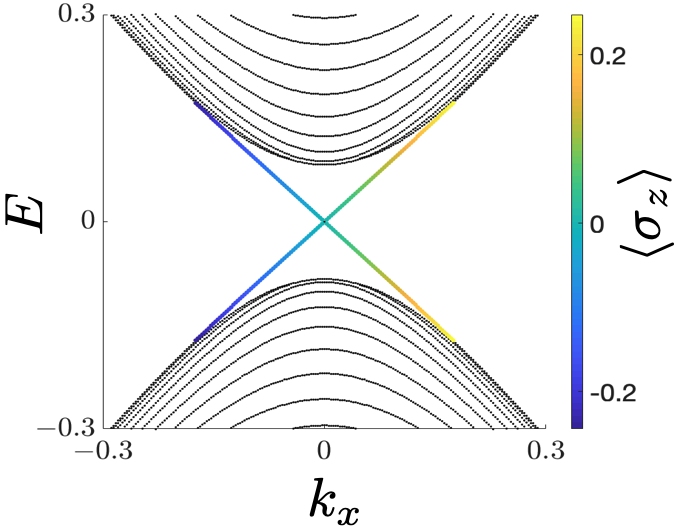,width=0.45\linewidth}}
\subfigure[]{\centering \epsfig{file=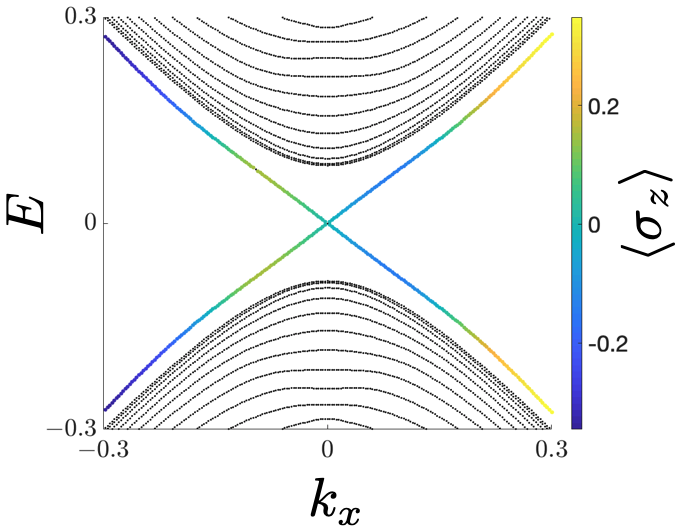,width=0.45\linewidth}}
\caption{(a) The bulk bands of $\hat{H}_0$ along the $k_z$ axis. The doubly degenerate bands cross at the two Dirac nodes at $\vk=\pm \vec{K}_0$. (b) The quasiparticle spectrum in the half BZ showing the two emergent Dirac nodes at $k_N=K_0+\sqrt{\mu^2+\Delta_0^2}$ and $k_S=K_0-\sqrt{\mu^2+\Delta_0^2}$.  (c) The bulk and surface quasiparticle energy spectra along the $k_x$ direction at $k_z=K_0+0.2$ with an open boundary at at $y=1$ and unit lattice constant. Bulk states are plotted in black and surface states are colored, with the color representing $\expect{\sigma_z}$. (d) Same as (c) with $k_z=K_0-0.2$. Parameter values are $L_y=100$, $m=t_1=t_2=1$, $\mu/t_1=0.2$, and $\Delta_0/t_1=0.2$. 
}
\label{fig: bandstructure}
\end{figure}

The matrix kernel in $\hat{H}_0$ is $h(\vk)=\sum_{j=1}^5 h_j(\vk)\Gamma_j$,
where $h_1(\vk)=m(2-\cos k_x-\cos k_y-\cos k_z)$, 
    $h_2(\vk)=t_1\sin k_x$, $h_3(\vk)=t_1\sin k_y$,
    $h_4(\vk)=t_2\sin k_x$, and $h_5(\vk)=t_2\sin k_y$.
Here $t_1$, $t_2$, and $m$ are hopping parameters which, for simplicity, satisfy $m=\sqrt{2}t_1=\sqrt{2}t_2$. 
The gamma matrices $\Gamma_i$ satisfying the Clifford algebra $\acomm{\Gamma_i}{\Gamma_j}=2\delta_{ij}s_0 \otimes \tau_0$ are chosen to be $\Gamma_1=s_0\otimes \tau_z$, $\Gamma_2=s_z \otimes \tau_x$, $\Gamma_3=s_0\otimes \tau_y$, $\Gamma_4=s_x\otimes\tau_x$, and $\Gamma_5=s_y\otimes\tau_x$, where $s_0$ ($\tau_0$) is the two-by-two identity matrix and $s_i$ ($\tau_i$) are the Pauli matrices in spin (orbital) space. There are two Dirac nodes at $\pm \vec{K}_0=\pm (0,0,K_0)^T$, where $K_0=\frac{\pi}{2}$, enclosed by the Fermi surfaces $\FS_{\pm}$ in the presence of doping, as illustrated by the bulk energy spectrum in Fig. \ref{fig: bandstructure} (a) along the $k_z$ axis.

The band Hamiltonian $\hat{H}_0$ has the symmetries required for realizing topological $\mathbb{Z}_2$ FSs. It preserves TRS, $\Theta h(\vk)\Theta^{-1}=h(-\vk)$, where  $\Theta=i s_y\otimes \tau_0 \circ K$ is the time-reversal operator satisfying $\Theta^2=-1$, and $K$ is the complex conjugation operator. Furthermore, since $h(\vk)$ is invariant under $k_z \rightarrow -k_z$, this symmetry can be considered as the `mirror' $M_z$,  although it keeps the spin and orbital spaces invariant. 
$\hat{H}_0$ also possesses 3D inversion symmetry, $\Pi h(\vk)\Pi^{-1}=h(-\vk)$, where $\Pi=\Gamma_1$. Combined with $M_z$, we have $P h(k_x,k_y,k_z)P^{-1}=h(-k_x,-k_y,k_z)$, where $P=\Pi M_z$. 
The eigenvalues of $P$ of the states on the FSs along the $\hat{\mathcal{T}}$-invariant line (the $k_z$ axis) are $\sgn h_1(0,0,k_z)=-\sgn (m\cos k_z)$, which changes sign at the Dirac nodes. Hence, the two Fermi surfaces $\FS_{\pm}$ are $\mathbb{Z}_2$ non-trivial by the Fu-Kane formula.

Now we consider the odd-parity pairing state $\bar{\Delta}(\vq)=i\Delta_0 s_y\otimes\tau_x$, which describes spin-singlet and inter-orbital pairing, as an example. The BdG quasiparticle energy spectrum along the $k_z$ axis is shown in Fig. \ref{fig: bandstructure} (b), which exhibits nodes at the poles of $\FS_+$. The full BdG Hamiltonian also possesses the symmetry $\mathcal{C}\equiv M_z\Xi$, where $\Xi=\nu_x\otimes s_0\otimes \tau_0\circ K$ is the charge conjugation operator and $\nu_i$ are the Pauli matrices in Nambu space. Along with the symmetry $\mathcal{T}$, each momentum slice labelled by $k_z$ can be regarded as a 2D TR-invariant topological superconductor belonging to the DIII class. Figs \ref{fig: bandstructure} (c) and (d) show two $k_z$ slices in the quasiparticle energy spectrum between the two emergent nodes with open boundary condition in the $y$-direction. Within the gap are helical surface states that are expected for a TR-invariant topological superconductor.

To illustrate the topological obstruction, we study the low-energy physics by projecting the pairing matrix onto the FSs. The $\hat{\mathcal{T}}$-Kramers doublet on $\FS_+$ can be chosen to be $\ket{\alpha_{+,1}(\vq)}=\left(u_\vq,v_\vq/\sqrt{2},0,v_\vq/\sqrt{2}\right)^T$ and  $\ket{\alpha_{+,2}(\vq)}=\left(0,v_\vq^*/\sqrt{2},u_\vq^*,-v_\vq^*/\sqrt{2} \right)^T$, where we choose $u_\vq=\cos\frac{\theta_\vq}{2}$ and $v_\vq=\sin\frac{\theta_\vq}{2}e^{-i\phi_\vq}$. The states satisfy the $\hat{\mathcal{T}}$-symmetry condition Eq. \eqref{eq: TR FS} and are regular over the entire FS except at the south pole, where the Dirac string lies.  Similarly, the eigenstates on $\FS_-$, which are related to $\ket{\alpha_{+,a}(\vq)}$ by $\hat{\Theta}$, are $\ket{\alpha_{-,1}(-\vq)}=\left(u_\vq,-v_\vq/\sqrt{2},0,-v_\vq/\sqrt{2} \right)^T$ and  $\ket{\alpha_{-,2}(-\vq)}=\left(0,-v_\vq^*/\sqrt{2},u_\vq^*,v_\vq^*/\sqrt{2} \right)^T$. Written in the above band basis, the projected pairing matrix is
\begin{equation}\label{eq: gap N}
    \Delta^N(\vq)=\frac{\Delta_0}{\sqrt{2}}\begin{pmatrix}
 -u_\vq^*v_\vq^*&\Rea u_\vq v_\vq^*\\
\Rea u_\vq v_\vq^*& u_\vq v_\vq
\end{pmatrix}.
\end{equation}
Locally, the pairing at a constant $k_z$ corresponds to that of a 2D helical topological superconductor. This description is not accurate globally, however. To determine the local pairing matrix near the south pole, we perform the gauge transformation in Eq. \eqref{eq: gauge transformation} with transition matrix $M_+(\vq)=ie^{i\phi_\vq \sigma_z}$, whose $\SU(2)$ part, $G_+(\vq)=e^{i \phi_\vq\sigma_z}$, belongs to topological class (II). The pairing matrix in gauge $S$, $\Delta^S(\vq)$, takes the same form as $\Delta^N(\vq)$ in Eq. \eqref{eq: gap N} but with $u_\vq=\cos\frac{\theta_\vq}{2}e^{i\phi_\vq}$ and $v_\vq=\sin\frac{\theta_\vq}{2}$. In this gauge, the Dirac string passes through the north pole and this pairing matrix is an accurate local description in the vicinity of the south pole. 

The local expressions for the pairing matrix near the north and south poles satisfy $\Delta^S(\vq)=\sigma_y \Delta^N(\vq) \sigma_y^T$. Therefore, a 2D momentum space slice labelled by $k_z$ near the north and south poles are related by reversing the pseudospins along with a phase change. Figs. \ref{fig: bandstructure} (c) and (d), which are momentum cuts near the north and south poles, respectively, illustrate this. Within the bulk gap are topological surfaces states with their $\expect{\sigma_z}$ values shown. When we move from north to south, the pseudospins of the low-energy surface excitations are reversed.

The pairing matrix $\Delta(\vq)$ can be expressed succinctly using time-reversal related monopole harmonics. 
To describe the orbital partial-waves in terms of eigenfunctions of $L_z$, we first reorganize the spin channel triplet pairing in the eigenbasis of $S_z$ using the tuple $\tilde{\vecg{\sigma}}=(\sigma_+,\sigma_z,\sigma_-)$ with $\sigma_{\pm}\equiv \mp (\sigma_x \pm i\sigma_y)/\sqrt{2}$. 
Then, $\Delta^{N/S}_{ab}(\vq)=\tilde{\vec{d}}^{N/S}(\vq)\cdot (\tilde{\vecg{\sigma}}i\sigma_y)_{ab}$ where $\tilde{\vec{d}}$ is denoted as  $(d_+,d_z,d_-)^T$ with $d_{\pm}\equiv \mp (d_x \mp id_y)/\sqrt{2}$
and $\tilde{\vec{d}}$ rotates as $\tilde{\vec{d}}^S(\vq)=\tilde{R}_+(\vq)\tilde{\vec{d}}^N(\vq)$. 
Here $\tilde{R}_+(\vq)=e^{iJ_z 2\phi_\vq}$ with $J_z=\diag (1,0,-1)$. In other words, the components $d_+$, $d_z$, and $d_-$ transform as monopole harmonics with monopole charges $1,0$, and $-1$, respectively. 
In the above example, 
$\tilde{\vd}(\hat{\vq})=-\sqrt{2\pi/3}  \Delta_0\left[Y_{q=1;1,0}, \ (Y_{1,1}-Y_{1,-1})/\sqrt{2}, \ Y_{q=-1;1,0} \right]^T(\hat{\vq})$, 
where $d_+ \propto Y_{q= 1;10}(\vq)$ and $d_- \propto Y_{q= -1;10}(\vq) = Y^*_{q= 1;10}(\vq+\pi \hat{\vecg{\phi}}_{\vq}) $  are a time-reversal pair of monopole harmonics in momentum space with opposite monopole charges $q=\pm 1$ (see S.M. II for the convention of momentum space monopole harmonics used here). $d_z$ is described by the usual $p_x$-wave spherical harmonics. 

The local gap functions can also be obtained using a non-singular gauge. Let $\ket{\chi_{+,1}(\vq)}=u_\vq^*\ket{\alpha_{+,1}(\vq)}+v_\vq \ket{\alpha_{+,2}(\vq)}$ and $\ket{\chi_{+,2}(\vq)}=-v_\vq^*\ket{\alpha_{+,1}(\vq)}+u_\vq\ket{\alpha_{+,2}(\vq)}$, which are non-singular as they only consist of products of monopole harmonic functions with opposite monopole charges \cite{Wu1977}. This gauge transformation corresponds to the change of basis matrix $U_+(\vq)=e^{-i\hat{\vec{n}}_\vq\cdot \vecg{\sigma}\theta_\vq/2}$, where $\hat{\vec{n}}_\vq=(\sin\phi_\vq,-\cos\phi_\vq,0)^T$, and, using Eq. \eqref{eq: gap sewing}, we obtain for the pairing matrix $\Delta(\vq)=(\tilde{\vd}(\vq)\cdot\tilde{\vecg{\sigma}})w_+(\vq+\pi \hat{\vecg{\phi}}_\vq)$, where the $d$-vector
\begin{align}
\tilde{\vd}(\vq)=\frac{\Delta_0}{2}\sin\theta_\vq \begin{bmatrix}
	-e^{-i\phi_\vq}(\cos\theta_\vq-\cos\phi_\vq\sin\theta_\vq)\\
\sqrt{2}(\sin\theta_\vq+\cos\phi_\vq\cos\theta_\vq)\\
	e^{i\phi_\vq}(\cos\theta_\vq-\cos\phi_\vq\sin\theta_\vq)
\end{bmatrix}
\end{align}
and sewing matrix $w_+(\vq+\pi \hat{\vecg{\phi}}_\vq)=e^{-i\hat{\vec{n}}_\vq\cdot \vecg{\sigma}\theta_\vq}i\sigma_y$. The pairing matrix $\Delta(\vq)$ is the same near the two poles. However, there is a twist in the basis: In the vicinity of the north pole, $\hat{\chi}_{+,1}^\dagger(\vq) \simeq \hat{\alpha}^\dagger_{+,1}(\vq)$ and $\hat{\chi}_{+,2}^\dagger(\vq) \simeq \hat{\alpha}^\dagger_{+,2}(\vq)$, but near the south pole, $\hat{\chi}_{+,1}^\dagger(\vq)\simeq i \hat{\beta}^\dagger_{+,2}(\vq)$ and $\hat{\chi}_{+,2}^\dagger(\vq)\simeq -i \hat{\beta}^\dagger_{+,1}(\vq)$. Because of the reversal of the indices $1$ and $2$, the pseudospins near the south pole are opposite to those at the north pole.

We remark that this phase is fundamentally different from currently known TR invariant topological superconductors, whose order parameters are not obstructed. For example, consider a superconductor with order parameter $\tilde{\vec{d}}(\vq)=\frac{\Delta_0}{2}\sin\theta_\vq(-e^{-i\phi_\vq},\sqrt{2}\cos\phi_\vq,e^{i\phi_\vq})^T$. This is $\tilde{\vd}^N$ in our model and satisfies the same symmetries. The non-trivial topology for this nodal phase arises by considering individual 2D slices labelled by $k_z$ and calculating the Fu-Kane invariant for each slice \cite{Sato2017,Schnyder2017}. Our system is also topological in this sense, but it has the further topological property that the order parameter is not well-defined over the FS, leading to the aforementioned topological twist in the spins of the quasiparticle spectrum.

{\it Conclusion.} -- To conclude, we have studied a three-dimensional, TR symmetric nodal superconducting phase whose order parameter is topologically obstructed over the FS, preventing it from being defined globally. This arises when the Cooper pairing is in a triplet state and between two FSs with non-trivial $\mathbb{Z}_2$ invariants, such as those in a $\mathbb{Z}_2$ Dirac semimetal. When the $\hat{\mathcal{T}}$-symmetry condition is imposed, the pairing matrix must be described using two gauge patches and the transition function between the pairing matrices in the two gauge patches corresponds to a non-contractible $\SO(3)$ rotation of the $\vec{d}$-vector. As a result of the topological obstruction, the pseudospins of the surface states are opposite at the poles. The results were also discussed in the sewing matrix formalism, which selects a globally well-defined gauge at the expense of the $\hat{\mathcal{T}}$-symmetry condition.

{\it Acknowledgment.} --
C.S. and Y.L. are supported by the U.S. Department of Energy, Office of Basic Energy Sciences, Division of Materials Sciences and Engineering, Grant No. DE-SC0019331. This work was supported in part by the Alfred P. Sloan Research Fellowships.


%

\balancecolsandclearpage
\section{Supplemental Materials}
\twocolumngrid
  \pagenumbering{arabic}
  \renewcommand{\thepage}{S-\arabic{page}}
\setcounter{equation}{0}
\setcounter{figure}{0}
\setcounter{table}{0}
\setcounter{page}{1}
\makeatletter
\renewcommand{\theequation}{S\arabic{equation}}
\renewcommand{\thefigure}{S\arabic{figure}}

\section{I. SU($2$) Wilson Loop}
In this supplementary material, the equivalence between the two $\mathbb{Z}_2$ invariants defined in the main text is established. The first invariant, $\delta_0$, is the relative sign between the $\SU(2)$ transition matrices at points related by $\hat{\mathcal{T}}$-symmetry \cite{Lee2008},
\begin{equation}
    G(\vq)=\delta_0 G(\vq+\pi \hat{\vecg{\phi}}_\vq).
\end{equation}
The second is the Fu-Kane formula in Eq. \eqref{eq: Pfaffian} \cite{Fang2016},
\begin{equation}\label{eq: Fu-Kane}
\delta=\Pf \tilde{w}_+(\theta_\vq=0)\Pf \tilde{w}_+(\theta_\vq=\pi),
\end{equation}
which characterizes the inability to select the $\SU(2)$ part of $\hat{\mathcal{T}}$ to be $i\sigma_y$ globally. As discussed in the main text, the two Fermi surfaces belong to the same topological class, so we will henceforth focus on $\FS_{+}$ and omit the $+$ subscript for notational convenience. 

\begin{figure}[htpb]
\centering  {\epsfig{file=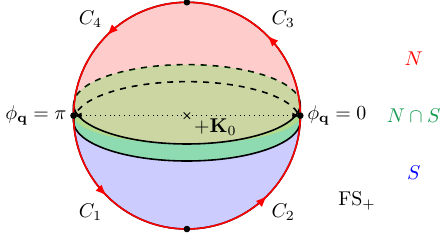,width=0.8\linewidth}}
\caption{The Fermi surface $\FS_+$ consists of two gauge patches, $N$ (red) and $S$ (blue), with overlap $N\cap S$ (green). The $\hat{\mathcal{T}}$-invariant Wilson loop $C$, which is separated into four segments $C_{1-4}$, is a great circle that passes through the north and south poles. For concreteness the meridians are taken to be at $\phi_\vq=0,\pi$.
}
\label{fig: Wilson}
\end{figure}

Following Ref. \cite{Lee2008}, the two $\mathbb{Z}_2$ invariants can be expressed as an $\SU(2)$ Wilson loop. In a gauge where the eigenstates are non-singular, $\ket{\chi_{a}(\vq)}\equiv \hat{\chi}_{a}^\dagger(\vq)\ket{0}$, it is possible to define the $\Un(2)$ Berry connection $\vec{A}_{ab}(\vq)=i\bra{\chi_{a}(\vq)}\nabla_\vq\ket{\chi_{b}(\vq)}$ over the entire FS. It is useful to decompose $\vec{A}$ into $\Un(1)$ and $\SU(2)$ parts: $\vec{A}=\vec{A}_{\Un(1)}+\vec{A}_{\SU(2)}$, where $\vec{A}_{\Un(1)}$ and $\vec{A}_{\SU(2)}$ are the traceful and traceless parts of $\vec{A}$, respectively. The central object connecting the two definitions for the $\mathbb{Z}_2$ invariant is the Wilson loop \cite{Lee2008,Fu2006}
\begin{equation}
    W[C]=\frac{1}{2}\Tr \mathcal{P}\exp\left[i\oint_C \d\vec{q}\cdot \vec{A}_{\SU(2)}(\vq)\right],
\end{equation}
where $\mathcal{P}$ is the path-ordering operator and $C$ is the $\hat{\mathcal{T}}$-invariant loop in Fig. \ref{fig: Wilson}, which is separated into four segments $C_{1-4}$. The Wilson loop is gauge invariant as a result of the trace.

To evaluate the Wilson loop, consider the unitary infinitesimal propagator \cite{Ryu2010}
\begin{equation}
    K_{ab}(\vq_2,\vq_1)\equiv\braket{\chi_a(\vq_2)}{\chi_b(\vq_1)}=[e^{i\d \vq\cdot \vec{A}(\vq_2)}]_{ab},
\end{equation}
where $\d \vq=\vq_2-\vq_1$. As a result of $\hat{\mathcal{T}}$-symmetry, it satisfies the sewing condition
    $K(\vq_1+\pi \hat{\vecg{\phi}}_\vq,\vq_2+\pi \hat{\vecg{\phi}}_\vq)=w(\vq_1)K^T(\vq_2,\vq_1)w^\dagger(\vq_2)$. When the propagator is decomposed into $\Un(1)$ and $\SU(2)$ parts, $K(\vq_1,\vq_2)=e^{i\d \vq\cdot \vec{A}_{\Un(1)}}\tilde{K}(\vq_1,\vq_2)$, where $\tilde{U}(\vq_2,\vq_1)=e^{i\d\vq\cdot \vec{A}_{\SU(2)}(\vq)}$, the $\SU(2)$ part of the propagator satisfies the corresponding sewing condition
    \begin{equation}\label{eq: prop sewing}
    \tilde{K}(\vq+\pi \hat{\vecg{\phi}}_\vq,\vq_2+\pi \hat{\vecg{\phi}}_\vq)=\tilde{w}(\vq_1)\tilde{K}^T(\vq_2,\vq_1)\tilde{w}^\dagger(\vq_2).
    \end{equation}
    Here $\tilde{w}(\vq)$ is the $\SU(2)$ part of the sewing matrix, as defined in the main text. The Wilson loop can be constructed from the infinitesimal propagators by
\begin{equation}
    W[C]=\frac{1}{2}\Tr \tilde{K}_4\tilde{K}_3\tilde{K}_2\tilde{K}_1,
\end{equation}
where the propagators for the four segments, $C_{1-4}$, are
\begin{align}
\begin{split}
    \tilde{K}_1&\equiv\prod_{n=1}^N \tilde{K}_\pi\left(\frac{\pi}{2}+n\delta\theta_\vq,\frac{\pi}{2}+(n-1)\delta\theta_\vq \right),\\
    \tilde{K}_2&\equiv\prod_{n=1}^N \tilde{K}_0\left(\pi-n\delta\theta_\vq,\pi-(n-1)\delta\theta_\vq\right),\\
    \tilde{K}_3&\equiv\prod_{n=1}^N \tilde{K}_0\left(\frac{\pi}{2}-n\delta\theta_\vq,\frac{\pi}{2}-(n-1)\delta\theta_\vq\right),\\
     \tilde{K}_4&\equiv\prod_{n=1}^N \tilde{K}_\pi\left(n\delta\theta_\vq,(n-1)\delta\theta_\vq \right).
    \end{split}
\end{align}
Here $\delta\theta_\vq=\frac{\pi}{2N}$ and we use the notation $\tilde{K}_{\phi_\vq}(\theta_{\vq_2},\theta_{\vq_1})=\tilde{K}(\vq_2,\vq_2)$, where $\vq_1=(\theta_{\vq_1},\phi_\vq)$ and $\vq_2=(\theta_{\vq_2},\phi_\vq)$. The  sewing condition gives the constraints, $\tilde{K}_1=\tilde{w}(\theta_\vq=\pi)\tilde{K}_2^T\tilde{w}^\dagger(\theta_\vq=\pi/2)$ and $\tilde{K}_4=\tilde{w}(\theta_\vq=\pi/2)\tilde{K}_3^T\tilde{w}^\dagger (\theta_\vq=0)$. Using $\tilde{w}(\theta_\vq=0,\pi)=\Pf \tilde{w}(\theta_\vq=0,\pi)i\sigma_y$ and the identity $\sigma_y F^T\sigma_y=F^\dagger$ for any $\SU(2)$ matrix $F$, the Wilson loop reduces to the Fu-Kane invariant, $W[C]=\delta$.

The Wilson loop $W[C]$ is also equal to the invariant $\delta_0$. To arrive at the appropriate form for $W[C]$, perform for the propagators in the segments $C_3$ and $C_4$ a gauge transformation to $\ket{\chi_a(\vq)}=\sum_b\ket{\alpha_b(\vq)}U_{ba}(\vq)$, where $\ket{\alpha_a(\vq)}\equiv \hat{\alpha}^\dagger_a(\vq)\ket{0}$ are states smooth over $N$ and satisfy the $\hat{\mathcal{T}}$-symmetry condition, Eq. \eqref{eq: TR FS}. Under this gauge transformation, the propagators transform as $\tilde{K}(\vq_2,\vq_1)=\tilde{U}^\dagger(\vq_2) \tilde{K}^N(\vq_2,\vq_1) \tilde{U}(\vq_1)$, where $\tilde{U}(\vq)$ and $\tilde{K}^N(\vq_2,\vq_1)$ are the $\SU(2)$ parts of $U(\vq)$ and $K^N_{ab}(\vq_2,\vq_1)\equiv \braket{\alpha_a(\vq)}{\alpha_b(\vq)}$, respectively. In this gauge, the $\hat{\mathcal{T}}$-symmetry condition is enforced, hence Eq. \eqref{eq: prop sewing} simplifies to $\tilde{K}^N(\vq_1+\pi \hat{\vecg{\phi}}_\vq,\vq_2+\pi \hat{\vecg{\phi}}_\vq)=[\tilde{K}^N(\vq_2,\vq_1)]^\dagger$. Consequently, the propagators $\tilde{K}^N$ in the entire segment $C_3\cup C_4$ cancel and $\tilde{K}_4\tilde{K}_3=\tilde{U}^\dagger(\theta_\vq=\pi/2,\phi_\vq=\pi) \tilde{U}(\theta_\vq=\pi/2,\phi_\vq=0)$. Similarly, the propagators in the segments $C_1$ and $C_2$ are evaluated in the gauge $S$ by performing the gauge transformation $\ket{\chi_a(\vq)}=\sum_{b}\ket{\beta_b(\vq)}V_{ba}(\vq)$, where $\ket{\beta_b(\vq)}\equiv \hat{\beta}^\dagger_{b}(\vq)\ket{0}$. By the same argument, only the gauge transformations at the end points contribute and $\tilde{K}_2\tilde{K}_1=\tilde{V}^\dagger(\theta_\vq=\pi/2,\phi_\vq=0) \tilde{V}(\theta_\vq=\pi/2,\phi_\vq=\pi)$. The change of basis matrices $U(\vq)$ and $V(\vq)$ are related to the transition matrix by $G(\vq)=\tilde{V}(\vq)\tilde{U}^\dagger(\vq)$. Hence, $W[C]=\frac{1}{2}\Tr[ G(\theta_\vq=\pi/2,\phi_\vq=\pi)G^\dagger(\theta_\vq=\pi/2,\phi_\vq=0)]=\delta_0$. This establishes that the two invariants, $\delta_0$ and $\delta$, are equal to the Wilson loop $W[C]$.

\section{II. Monopole harmonics in momentum space}
The monopole harmonics in momentum space are defined as irreducible representations of the gauge-covariant angular momentum operator 
	\begin{equation}
	\vec{L}(\vk)=-\hbar\vk\times (i\nabla_\vk -\vec{A}(\vk))+\hbar q\hat{\vk},
	\end{equation}
	where the $\Un(1)$ Berry connection of the state $\ket{u(\vk)}$, $\vec{A}(\vk)=\bra{u(\vk)}i\nabla_\vk \ket{u(\vk)}$, describes a magnetic monopole with monopole charge $q$ at the origin in momentum space. More concretely, we take the Berry connections in the regions $N$ and $S$ to be
	\begin{align}
		\vec{A}^N&=q\frac{1-\cos\theta_\vk}{k\sin\theta_\vk}\hat{\vecg{\phi}}_\vk,&\vec{A}^S&=-q\frac{1+\cos\theta_\vk}{k\sin\theta_\vk}\hat{\vecg{\phi}}_\vk.
	\end{align}
	Note that our definition of the angular momentum operator in momentum space differs from Ref. \onlinecite{Wu1976}, which is defined in real space as $\vec{L}(\vr)=\vr \times (-i\hbar \nabla_\vr - \vec{A}(\vr))-\hbar q \hat{\vr}$, by a minus sign in the monopole charge $q$, and hence our monopole harmonic $Y_{q;lm}(\hat{\vk})$ has the same functional form as $Y_{-q;lm}(\hat{\vr})$ in Ref. \onlinecite{Wu1976}. 
	
	The monopole harmonics can be written in terms of the Wigner $D$-matrices as
		\begin{align}
			Y_{q;lm}(\theta_\vk,\phi_\vk,\psi_\vk)&=\sqrt{\frac{2l+1}{4\pi}}D_{m,q}^{(l)*}(\phi_\vk,\theta_\vk,\psi_\vk)\\
			&=\sqrt{\frac{2l+1}{4\pi}}d_{m,q}^{(l)*}(\theta_\vk)e^{im\phi_\vk}e^{iq\psi_\vk},
		\end{align}
		where the Wigner matrices are defined as 
		\begin{align}
		\label{eq: Wigner D}
		D^{(l)}_{mn}(\phi_\vk,\theta_\vk,\psi_\vk)&=\bra{l,m}e^{-\frac{i}{\hbar} \hat{J}_z \phi_\vk}e^{-\frac{i}{\hbar}\hat{J}_y \theta_\vk}e^{-\frac{i}{\hbar}\hat{J}_z\psi_\vk}\ket{l,n}\nonumber\\
		d^{(l)}_{mn}(\theta_\vk)&=\bra{l,m}e^{-\frac{i}{\hbar}\hat{J}_y\theta_\vk}\ket{l,n}.
		\end{align}
	Here $(\phi_\vk,\theta_\vk,\psi_\vk)$ are the usual Euler angles. The choice of the third Euler angle, $\psi_\vk$, corresponds to the choice of gauge. For example, the vector potentials $\vec{A}^N$ and $\vec{A}^S$ correspond to the gauge choices $\psi_\vq=-\phi_\vq$ and $\psi_\vq=\phi_\vq$, respectively. Examples of monopole harmonics in the gauge $N$ with $q=0,\pm 1$ and $l=1$, including those used in the main text, are listed in Tab. \ref{tab: monopole harmonics}. Those in the gauge $S$ can be obtained by the gauge transformation $Y^N_{q;lm}(\hat{\vk})=Y^S_{q;lm}(\hat{\vk})e^{-2iq\phi_\vk}$.
	
	\begin{table}[htpb]
	    \centering
	    \begin{tabular}{|c|ccc|}
	    \hline
	        \diagbox{$m$}{$q$} & $1$&$0$& $-1$\\
	   \hline
	         $1$&$\frac{1}{2}(1+\cos\theta_\vk)$ &$-\frac{1}{\sqrt{2}}\sin\theta_\vk e^{i\phi_\vk}$&$\frac{1}{2}(1-\cos\theta_\vk)e^{i2\phi_\vk}$\\
	         $0$&$\frac{1}{\sqrt{2}}\sin\theta_\vk e^{-i\phi_\vk}$&$\cos\theta_\vk$&$-\frac{1}{\sqrt{2}}\sin\theta_\vk e^{i\phi_\vk}$\\
	         $-1$&$\frac{1}{2}(1-\cos\theta_\vk)e^{-i2\phi_\vk}$&$\frac{1}{\sqrt{2}}\sin\theta_\vk e^{-i\phi_\vk}$&$\frac{1}{2}(1+\cos\theta_\vk)$ \\
	         \hline
	    \end{tabular}
	    \caption{The monopole harmonics $\sqrt{\frac{4\pi}{3}}Y_{q;lm}(\hat{\vk})$ with $l=1$ and $q,m=0,\pm 1$ in the gauge $N$.}
	    \label{tab: monopole harmonics}
	\end{table}
\end{document}